\documentclass[aps,prx,twocolumn,preprintnumbers,superscriptaddress,10pt,showkeys,floatfix]{revtex4-2}
\usepackage{amsmath}
\usepackage{graphicx}
\usepackage{dcolumn}
\usepackage{bm}
\usepackage{hyperref}
\usepackage{colortbl}
\usepackage{times}

\definecolor{ao}{rgb}{0.0, 0.5, 0.0}
\newcommand{\dd}{\mathrm{d}}

\definecolor{ao}{rgb}{0.0, 0.5, 0.0}

\definecolor{ktcolor}{rgb}{0.725, 0.031, 0.886}

\newcommand{\soft}{\texttt{Soft}}
\newcommand{\softnpt}{\texttt{Soft-NPT}}
\newcommand{\inter}{\texttt{Inter}}
\newcommand{\stiff}{\texttt{Stiff}}
\newcommand{\Yq}{$\langle Y_{\rm quark}\rangle$}

\definecolor{mjc}{rgb}{1.0, 0.5, 0.0}

\begin{document}

\title{Quark formation and phenomenology in binary neutron-star mergers
  using V-QCD}

\author{Samuel Tootle}
\author{Christian Ecker}
\author{Konrad Topolski}
\affiliation{Institut f\"ur Theoretische Physik, Goethe Universit\"at, Max-von-Laue-Str. 1, 60438 Frankfurt am Main, Germany}
\author{Tuna Demircik}
\affiliation{Asia Pacific Center for Theoretical Physics, Pohang, 37673, Korea}
\author{Matti J\"arvinen}
\affiliation{Asia Pacific Center for Theoretical Physics, Pohang, 37673, Korea}
\affiliation{Department of Physics, Pohang University of Science and Technology, Pohang, 37673, Korea}
\author{Luciano Rezzolla}
\affiliation{Institut f\"ur Theoretische Physik, Goethe Universit\"at, Max-von-Laue-Str. 1, 60438 Frankfurt am Main, Germany}
\affiliation{Frankfurt Institute for Advanced Studies, Ruth-Moufang-Str. 1,
60438 Frankfurt, Germany}
\affiliation{School of Mathematics, Trinity College, Dublin 2, Ireland}

\date{\today}
 
\begin{abstract}
Using full 3+1 dimensional general-relativistic hydrodynamic simulations
of equal- and unequal-mass neutron-star binaries with properties that are
consistent with those inferred from the inspiral of GW170817, we perform
a detailed study of the quark-formation processes that could take place
after merger. We use three equations of state consistent with current
pulsar observations derived from a novel finite-temperature framework
based on V-QCD, a non-perturbative gauge/gravity model for Quantum
Chromodynamics. In this way, we identify three different post-merger
stages at which mixed baryonic and quark matter, as well as pure quark
matter, are generated. A phase transition triggered collapse already
$\lesssim 10\,\rm{ms}$ after the merger reveals that the softest version
of our equations of state is actually inconsistent with the expected
second-long post-merger lifetime of GW170817. Our results underline the
impact that multi-messenger observations of binary neutron-star mergers
can have in constraining the equation of state of nuclear matter,
especially in its most extreme regimes.
\end{abstract}

\keywords{neutron star mergers, critical point, gauge/gravity duality}

\preprint{APCTP Pre2022 - 007}

\maketitle

\section{Introduction\label{sec:Intro}}

Multi-messenger observations from binary neutron-star mergers are
expected to provide new insights into the properties of dense and hot
Quantum Chromodynamics (QCD). However, due to the lack of first principle
techniques, the precise phase structure of QCD and its equation of state
(EOS) are currently not known at densities and temperatures that occur
during and after the merger of two neutron stars. It is, therefore, not
clear if a transition from dense nuclear to quark matter can happen in
such events nor if such a transition leaves a discernible imprint on the
gravitational waveform or the electromagnetic counterpart. Such an
imprint may be best visible in the kHz gravitational-wave signal emitted
during a possible post-merger hypermassive neutron star (HMNS) stage as
it encodes information on the hot and dense part of the EOS that is
inaccessible during the inspiral phase, however, this parameter space
will only be able to be reliably explored by future detectors. In order
to resolve the question of the detectability of quark matter in binary
neutron-star mergers, it is therefore crucial to explore state-of-the-art
models that make predictions for the phase structure of QCD beyond the
saturation density of atomic nuclei $n_s=0.16\,\rm fm^{-3}$ and, at the
same time, satisfy known constraints from theory and astrophysical
observations.

Numerical simulations of binary neutron-star systems play a critical role
to ascertain the influence of the neutron-star characteristics (mass and
spin) and the EOS on the post-merger remnant and lifetime
\cite{Baiotti2016,Paschalidis2016}. For EOSs that include a transition to
quark matter, these simulations also provide insights into the mechanisms
for and the abundance of quark formation in the post-merger remnant.
There exists a number of works that study mergers using models with a
phase transition either in an effective particle
approach~\cite{Bauswein:2018bma} or in fully general-relativistic
hydrodynamics~\cite{Most2018b,Most2019c,Ecker:2019xrw}. The recent
study~\cite{Prakash:2021wpz} also analyses the composition of the merger
ejecta. In~\cite{Weih2020} possible signatures of phase transitions in
the gravitational wave signal have been classified using a polytropic EOS
with the phase transition to quark matter introduced via a Gibbs-like
construction and temperature dependence using the standard $\Gamma_{\rm
  th}$ component based on an ideal-fluid EOS~\cite{Rezzolla_book:2013}.

In this article we perform the first study to analyse quark matter
production in BNS mergers using a novel framework~\cite{Demircik:2021zll}
for describing nuclear matter which combines the gauge/gravity duality
with nuclear theory models. In this combined framework, the EOS at large
densities is given by the gauge/gravity
duality~\cite{Maldacena:1997re,Witten:1998qj}, more precisely the V-QCD
model~\cite{Jarvinen:2011qe,Ishii:2019gta,Ecker2019}, which is used to
describe the deconfinement phase transition from dense baryonic to quark
matter. The DD2 version of the Hempel-Schaffner-Bielich statistical
model~\cite{Typel:2009sy,Hempel:2009mc} is used at low densities, and a
van der Waals model for the temperature dependence in the dense nuclear
matter phase. The advantage of this combined framework is that it is, by
construction, consistent with theoretic predictions from nuclear theory
at low densities and perturbative QCD at high densities. At the same
time, it provides a description for the deconfinement phase transition
within a single model, i.e., V-QCD. The framework also yields predictions
for the location of the QCD critical point which allows us to investigate
the possibility that neutron star matter during and/or after the merger
reach temperatures sufficient to probe this part of the QCD phase
diagram.

\begin{figure*}[htb]
    \includegraphics[height=0.34\textwidth]{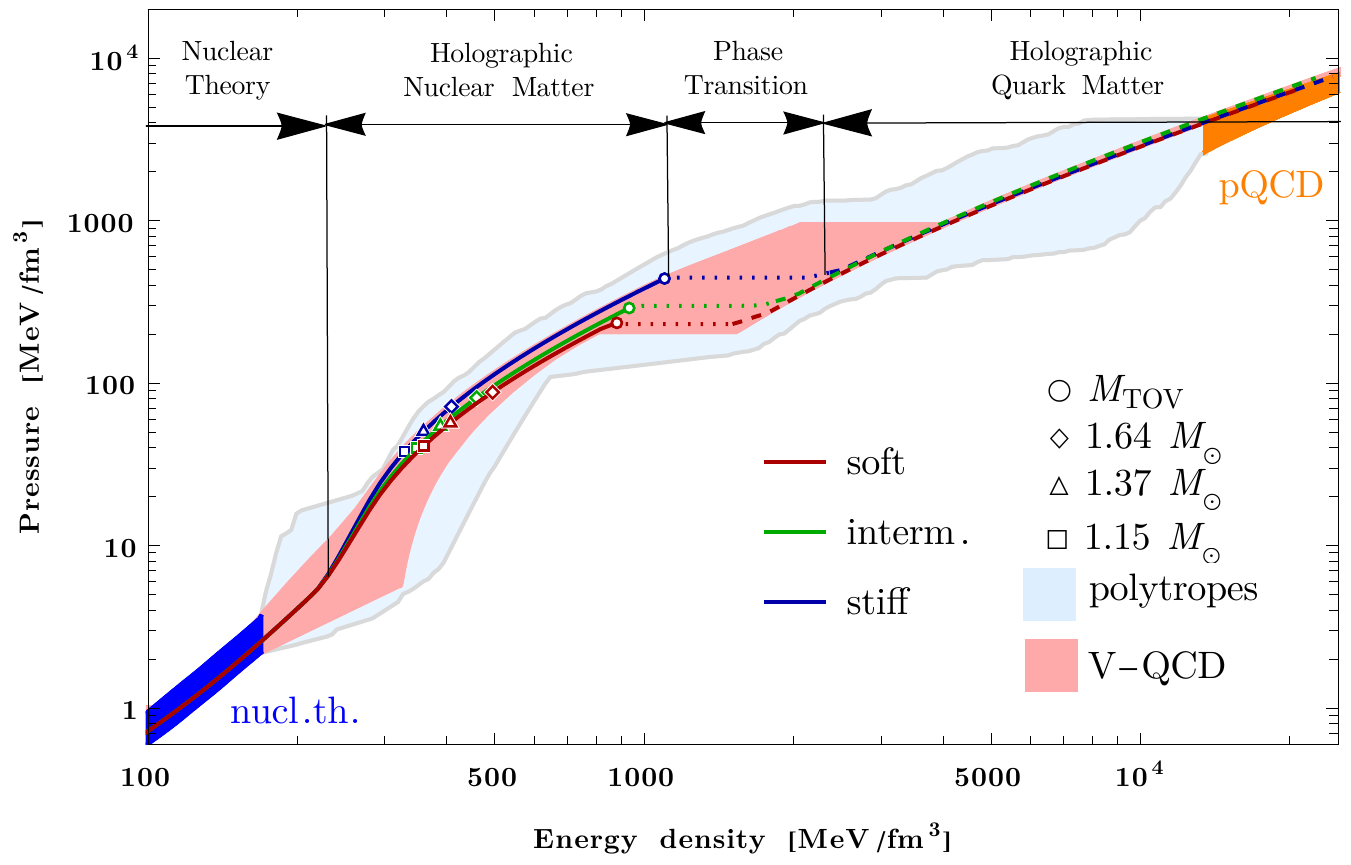}\quad
    \includegraphics[height=0.34\textwidth]{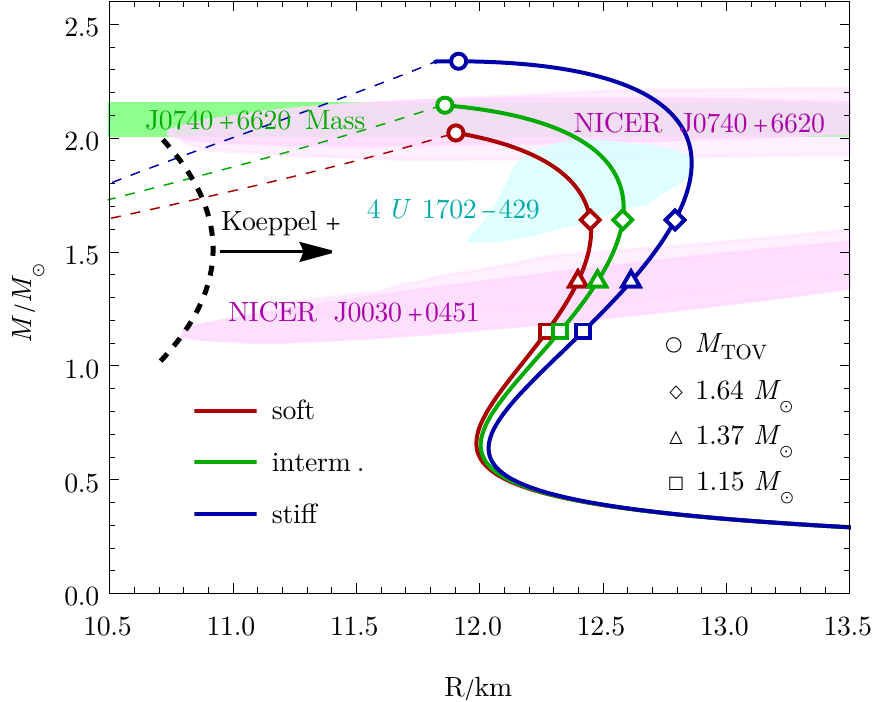}
    \caption{\small Left: Cold, beta equilibrium slices of the V-QCD
      EOS. Shown are the \soft\ (red), \inter\ (green) and \stiff\ (blue)
      versions of the model. Dark blue and orange bands mark the
      uncertainties in nuclear theory \cite{Hebeler:2013nza} and
      perturbative QCD \cite{Fraga:2013qra} calculations, respectively.
      The light blue region marks the theoretically allowed region
      spanned by polytropic EOSs, where the pink region marks the
      parametric freedom of V-QCD after imposing constraints from NS
      observations. Right: Corresponding mass-radius relations where
      dashed lines mark unstable quark matter branches. In addition we
      show various error bands of direct mass~\cite{NANOGrav:2019jur,
        Fonseca:2021wxt} and radius~\cite{Riley:2019yda, Miller:2019cac,
        Nattila:2017wtj, Miller:2021qha, Riley:2021pdl} measurements of
      heavy pulsars; shown with a black dashed line is a theoretical
      lower bound for the radii as computed from considerations on the
      threshold mass by Koeppel et al.~\cite{Koeppel2019}. }
    \label{fig:MR}
\end{figure*}

When taking into account different individual masses and spins of the
stars as well as variations of the EOS, the parameter space of binary
neutron star simulations grows rapidly. As a first test of the novel
V-QCD EOS, we study configurations motivated by information gained from
the inspiral part of the gravitational-wave event
GW170817~\cite{LIGOScientific:2017vwq} in order to determine whether or
not the constructed EOSs agree with the observation. To do so, we perform
full 3+1 dimensional general-relativistic hydrodynamics (GRHD)
simulations consistent with GW170817 so as to determine mechanisms for
quark production as well as ascertain the importance of multi-messenger
detections on constraining the model. More specifically, we simulate
non-spinning binaries of fixed chirp mass $\mathcal{M}_{\rm chirp} :=
{(M_1 M_2)^{3/5}}/{(M_1+M_2)^{1/5}} = 1.186\,M_\odot$ and two different
mass ratios $q:=M_2/M_1=1,\,0.7$. We use three different EOSs, based on a
soft, an intermediate and a stiff version of V-QCD, whose cold
beta-equilibrium parts
satisfy~\cite{Jokela:2020piw,Demircik:2020jkc,Jokela:2021vwy} all
currently known observational constraints from direct
mass~\cite{Antoniadis:2013pzd,NANOGrav:2019jur,Fonseca:2021wxt} and
radius~\cite{Miller:2021qha,Riley:2021pdl,Riley:2019yda,Miller:2019cac}
measurements of heavy pulsars as well as the constraint on the tidal
deformability obtained from GW170817~\cite{LIGOScientific:2018cki}.

As an important outcome of our simulations we identify three different
stages of quark matter formation during the post-merger HMNS evolution
which will be referenced as hot-quark (HQ), warm-quark (WQ) and
cold-quark (CQ) stages. The basis and definition of these stages are
discussed in detail in Section~\ref{sec:dynamics}. Furthermore, we
provide an analysis of the gravitational waveforms and their frequency
spectra of all our simulations. The clearest imprint on the waveform is
their early termination caused by the phase-transition-triggered collapse
to a black hole in the CQ stage. From their frequency spectra we extract
a number of characteristic post-merger frequencies.

The rest of the article is structured as follows. In
Section~\ref{sec:Model} we provide more details about the V-QCD model.
In Section~\ref{sec:id:evolution} we explain our numeric setup for the
binary initial data and the subsequent time evolution. In
Section~\ref{sec:dynamics} we present our results for the merger dynamics
and show examples for the three post-merger stages mentioned earlier. In
Section~\ref{sec:gwspectra} we discuss our results for the gravitational
waves and their spectra. In Section~\ref{sec:Conclusion} we summarise and
conclude. Finally, in Appendix~\ref{sec:App1:conv} we show results
obtained with lower resolution as in the main text and study the
consistency of our simulations. Hereafter, we use units where we set
$k_{_{\rm B}} = G = c = 1$ unless otherwise noted.

\section{Matter Model}
\label{sec:Model}

\begin{table*}[th]
 \begin{ruledtabular}
 \begin{tabular}{l|ccccccccccccccc}
 
   & $q$ & ${M_{\rm TOV}}$ & ${M_1}$ & ${M_2}$ & ${R_1}$ & ${R_2}$ & $\Lambda_1$ & $\Lambda_2$ & $\tilde{\Lambda}$& $\Lambda_{1.4}$ & ${f_{\rm mer}}$ & ${f_{3}}$ & ${f^{2,1}}$ & ${f^{2,2}}$ & ${t_{\rm BH}}$  \\
   &  & $[M_\odot]$ & $[M_\odot]$ & $[M_\odot]$ & $[{\rm km}]$ & $[{\rm km}]$ & & & & & ${\rm kHz}$ & ${\rm kHz}$ & ${\rm kHz}$ & ${\rm kHz}$ & $[{\rm ms}]$  \\
  \hline
  \texttt{Soft\_q10~~~~}    & $1.0$ & $2.02$ & $1.37$ & $1.37$ & $12.37$ & $12.37$ & $537$ & $~537$ & $537$ & $475$ & $1.77$ & $4.00$ & $1.44$ & $2.92$ & $9.5$
  \\  
  \texttt{Soft\_q07~~}    & $0.7$ & $2.02$ & $1.64$ & $1.15$ & $12.42$ & $12.24$ & $183$ & $1393$ & $517$ & $475$ & $1.63$ & $3.80$ & $1.55$ & $2.80$ & $5.8$
  \\  
  \texttt{Soft\_q10-NPT}    & $1.0$ & $2.06$ & $1.37$ & $1.37$ & $12.37$ & $12.37$ & $537$ & $~537$ & $537$ & $475$ & $1.76$ & $4.00$ & $1.62$ & $2.85$ & $>37$
  \\  
  \texttt{Soft\_q07-NPT}  & $1.0$ & $2.06$ & $1.64$ & $1.15$ & $12.42$ & $12.24$ & $183$ & $1393$ & $517$ & $475$ & $1.64$ & $3.85$ & $1.47$ & $2.79$ & $11$
  \\
  \texttt{Inter\_q10~~~}    & $1.0$ & $2.14$ & $1.37$ & $1.37$ & $12.45$ & $12.45$ & $565$ & $~565$ & $565$ & $511$ & $1.74$ & $4.00$ & $1.51$ & $2.73$ & $>35$
  \\  
  \texttt{Inter\_q07~}    & $0.7$ & $2.14$ & $1.64$ & $1.15$ & $12.56$ & $12.30$ & $201$ & $1437$ & $543$ & $511$ & $1.63$ & $3.70$ & $1.42$ & $2.67$ & $>37$
  \\  
  \texttt{Stiff\_q10~~~}    & $1.0$ & $2.34$ & $1.37$ & $1.37$ & $12.58$ & $12.58$ & $617$ & $~617$ & $617$ & $560$ & $1.74$ & $3.90$ & $1.39$ & $2.49$ & $>37$
  \\  
  \texttt{Stiff\_q07~}    & $0.7$ & $2.34$ & $1.64$ & $1.15$ & $12.76$ & $12.38$ & $231$ & $1525$ & $591$ & $560$ & $1.59$ & $3.50$ & $1.39$ & $2.48$ & $>38$
  \\
\end{tabular}
\end{ruledtabular}
 \caption{Properties of cold non-spinning isolated neutron stars used in
   the construction of the binary initial data.  Specifically, for each
   model considered, we list the mass ratio, $q$; the maximum non-rotating mass, $M_{\rm TOV}$;
   the gravitational mass of each object at infinite separation, $M_{<1/2>}$;
   the proper radius of each object, $R_{<1/2>}$; the tidal deformability of each object, $\Lambda_{<1/2>}$;
   the binary tidal deformability, $\tilde{\Lambda}$; the tidal deformability of a $1.4~M_\odot$ NS, $\Lambda_{1.4}$;
   the instantaneous frequency at the maximum GW strain amplitude, $f_{\rm mer}$;
   the characteristic frequencies of the post-merger phase, $f_3 \,, f^{2,1} \,, f^{2,2}$;
   and the collapse time to black hole formation, $t_{\rm BH}$, for the configurations that collapsed within the evolution time.
   }
\label{tab:NSprop} 
\end{table*}

In this section we provide a brief summary of the combined framework used
in our merger simulations. The most important novel ingredient in this
framework is the use of the gauge/gravity duality. The gauge/gravity
duality allows one to translate intractable problems in quantum field
theory at strong coupling and finite density to tractable problems in
classical five-dimensional gravity. This approach has been successfully
applied in the context of heavy-ion
collisions~\cite{Casalderrey-Solana:2011dxg,Busza:2018rrf}, condensed
matter physics~\cite{Zaanen:2015oix,Hartnoll:2016apf} and, more recently,
also astrophysics~\cite{Hoyos:2016zke, BitaghsirFadafan:2019ofb,
  Jarvinen:2021jbd, Hoyos:2021uff, Kovensky:2021kzl, Bartolini:2022rkl}.
The specific gauge/gravity model which we are using is V-QCD, which we
will now briefly review, however, a detailed introduction to V-QCD and to
the EOS at finite temperature and charge fraction based on V-QCD can be
found in~\cite{Jarvinen:2021jbd,Demircik:2021zll}.

V-QCD includes sectors both for
quarks~\cite{Bigazzi:2005md,Casero:2007ae} and for
gluons~\cite{Gursoy:2007cb,Gursoy:2007er}. It is based on a string
theory setup in the Veneziano
limit~\cite{Veneziano:1976wm,Veneziano:1979ec} in which both the number
of colors $N_c$ and flavors $N_f$ are large, but their ratio is
$\mathcal{O}(1)$~\cite{Jarvinen:2011qe,Alho:2012mh,Alho:2013hsa} such
that the quark degrees of freedom strongly back-react to the gluon
dynamics. However, the model cannot be strictly derived from string
theory, and one switches to an effective ``bottom-up'' approach where the
action of the model is tuned to agree with QCD data at finite $N_c$ and
$N_f$. Therefore the model includes a relatively large number of
parameters that are tuned to reproduce known features of QCD such as
asymptotic freedom, confinement, linear glueball and meson trajectories
and chiral symmetry breaking. The remaining freedom is constrained by
lattice data for large-$N_c$ pure Yang-Mills theory~\cite{Panero:2009tv}
and data for $N_c=3$ QCD with $N_f=2+1$ flavors at physical quark
masses~\cite{Alho:2015zua,Jokela:2018ers} at small baryon-number
density. Such an effective approach turns out to be useful to model dense
nuclear and quark matter in the region where the QCD-coupling is large
and traditional nuclear theory calculations and perturbation theory are
not reliable~\cite{Jokela:2018ers,Hoyos:2020hmq,Hoyos:2021njg}.

In the current version of the model baryons are implemented in a
homogeneous approximation~\cite{Ishii:2019gta}. This approximation is
natural at densities well above $n_s$, where the distance between
neighbouring nucleons becomes smaller than their diameter and a
description in terms of homogeneous matter is expected to work well. At
densities below $n_s$ this homogeneous approximation breaks down and
traditional nuclear matter models become more reliable. In the combined
framework this is taken into account by constructing hybrid EOSs whose
cold low-density part is modelled by nuclear theory and the dense
baryonic and quark matter part via the gauge/gravity duality. For the
cold nuclear theory part we use a combination~\footnote{By using the 
combination of the HS(DD2) and APR models together with V-QCD, we obtain 
EoSs that agree well with all observational constraints. 
Using only the HS(DD2) model would lead to a slight tension with 
the measurement of the tidal deformability from GW170817.} of the
Hempel-Schaffner-Bielich (HS) EOS~\cite{Hempel:2009mc} with DD2
relativistic mean field theory interactions~\cite{Typel:2009sy} and the
Akmal-Pandharipande-Ravenhall (APR) model~\cite{Akmal:1998cf}.

The prediction of V-QCD for the EOS in the nuclear matter phase suffers
from a generic limitation of gauge/gravity duality: the temperature
dependence is trivial. In \cite{Demircik:2021zll} this issue was solved
by using a simple van der Waals model, i.e., gas of nucleons, electrons
and mesons with excluded volume correction for the nucleons and an
effective potential. The van der Waals model was chosen to match with the
V-QCD prediction for the cold EOS of nuclear matter and then used to
extrapolate the result to finite temperature. The model was further
improved to allow deviation from beta-equilibrium by imposing the charge
fraction dependence of the HS(DD2) model
\cite{Hempel:2009mc,Typel:2009sy} for nuclear matter and a simple model
of free electrons for quark matter. In the regime of low density, both
the dependence on the temperature and the charge fraction are following
the HS(DD2) model. Finally, our framework contains a strong first order
phase transition between the nuclear and quark matter phases at low
temperatures which ends on a critical point at finite temperature and
density. Consequently, there is a mixed nuclear-quark matter phase which
was obtained in~\cite{Demircik:2021zll} by carrying out a Gibbs
construction depending on all the three variables, i.e., the density, the
temperature, and the charge fraction. Surface tension of the nuclear to
quark matter interface is neglected. This is expected to be a good
approximation at the temperatures that we encounter in our simulations,
because the latent heat at the transition is very high, comparable to the
free energy density of the nuclear matter phase.

We here use three variants of the EOS, which were established
in~\cite{Demircik:2021zll} in order to represent the parameter dependence
of the model. These are called the soft (\soft), intermediate (\inter),
and stiff (\stiff) variants; and reflect three different choices of the
parameters in the action of V-QCD~\cite{Jokela:2018ers}. In
Fig.~\ref{fig:MR} (left) we show cold beta-equilibrium slices of these
three EOSs with uncertainty bands from nuclear
theory~\cite{Hebeler:2013nza} (blue) and perturbative
QCD~\cite{Fraga:2013qra} (orange). Red, green and blue lines are the
\soft, \inter\ and \stiff\ versions of V-QCD where the dotted part of
these curves mark the first order phase transition between the baryonic
(solid) the quark matter (dashed) phase. In addition we show markers for
the central densities reached in various isolated non-rotating stars that
we use to initialise the binary systems in our simulations. Light blue
and pink regions mark the residual freedom of polytropic parametrizations
of the EOS~\cite{Annala:2019puf,Annala:2021gom} and of the V-QCD
model~\cite{Jokela:2020piw}, respectively, after imposing constraints
from neutron star observations. In Fig.~\ref{fig:MR} (right) we show the
corresponding mass-radius relations of non-rotating stars together with
various error bands of the direct mass measurement (green area) of the
heaviest known pulsar
PSR~J0740+6620~\cite{NANOGrav:2019jur,Fonseca:2021wxt} ($M=2.08 \pm 0.07
M_\odot$) and direct radius measurements of
PSR~J0740+6620~\cite{Miller:2021qha,Riley:2021pdl} and
PSR~J0030+0451~\cite{Riley:2019yda,Miller:2019cac} obtained by the NICER
experiment (pink ellipses) as well as from the measurement of the X-ray
binary 4U~1702-429~\cite{Nattila:2017wtj} (cyan area;
see~\cite{Jokela:2021vwy} for a more detailed analysis of the impact of
the NICER results). Finally, shown with a black dashed line is a
theoretical lower bound for the radii as computed from considerations on
the threshold mass by Koeppel et al.~\cite{Koeppel2019}. We note that we
have checked that our EOSs respect the upper bound on the binary tidal
deformability $\tilde \Lambda<720$ (low-spin priors) obtained from
GW171817 and are well within the bounds recently proposed on general
parameterization of the sound speed in neutron stars
\cite{Altiparmak:2022} (see Table~\ref{tab:NSprop}).

\section{Initial Data and Evolution Setup}
\label{sec:id:evolution}

To generate the binary neutron-star initial data we use the recently
developed Frankfurt University/Kadath
(\texttt{FUKA})~\citep{Papenfort2021b} solver library. \texttt{FUKA} is
based on an extended version of the \texttt{KADATH} spectral solver
library~\citep{Grandclement09}, which has been specifically designed for
numerical relativity applications. To generate the initial data for
compact binaries, \texttt{FUKA} uses the eXtended Conformal Thin Sandwich
formulation of Einstein's field equations~\citep{Pfeiffer:2002iy,
  Pfeiffer:2005, Papenfort2021b}. Initially \texttt{FUKA} solves the
initial data using force-balance equations to obtain a binary in the
quasi-circular orbit approximation. In an effort to minimize the
residual eccentricity during the inspiral, an eccentricity reduction step
is performed using estimates for the orbital- and radial-infall-
velocities at $3.5$-th post-Newtonian order. As shown
in~\citep{Papenfort2021b}, this procedure leads to a significant
reduction of the eccentricity in asymmetric and spinning compact object
binaries. We find the advanced handling of eccentricities in
\texttt{FUKA} essential to obtain accurate initial data for the unequal
mass binaries studied in this work.

\begin{figure*}[ht!]
 \includegraphics[width=0.9\textwidth, keepaspectratio]{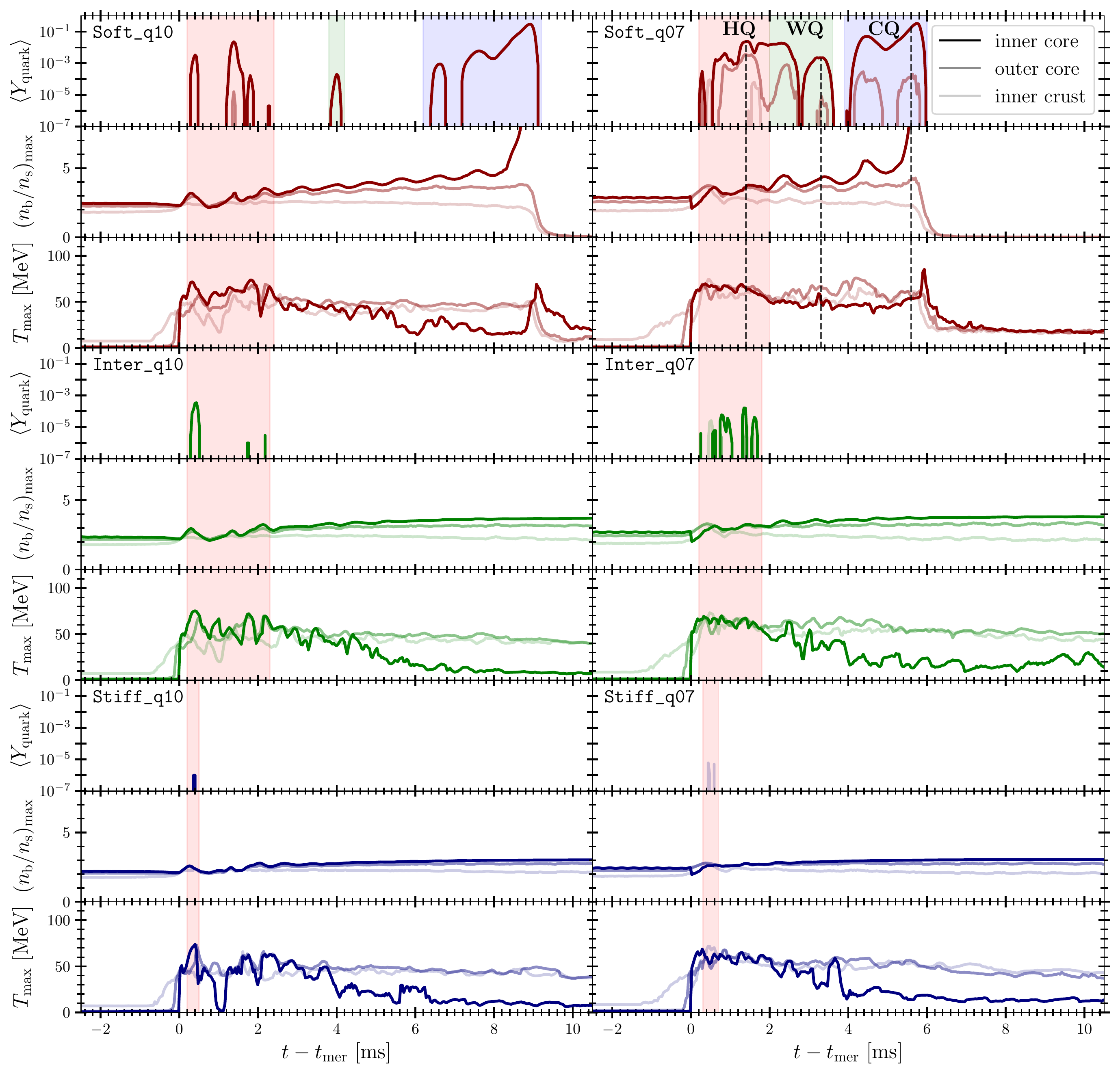}
 \caption{Correlation between the maximal temperature, maximal density
   and quark fraction in the three regions of interest, for all three
   equations of state and both mass ratios. We mark the time slices used
   for Fig. \ref{fig:soft_q0.7_hr_3x3_plot2D} with vertical black
   lines. The abbreviations HQ, WQ and CQ
   stand for the hot-quark, warm-quark and the cold-quark stages
   respectively.}
 \label{fig:yq_rho_temp_correl_annuli_combined}
\end{figure*}

For the binary evolution we make use of the \texttt{Einstein
  Toolkit}~\cite{EinsteinToolkit_etal:2020_11} infrastructure that
includes the fixed-mesh box-in-box refinement framework \texttt{Carpet}
\cite{Schnetter-etal-03b}. For the simulations presented in the main text
we use six refinement levels with finest grid-spacing of
$\Delta_\texttt{H} := 221\,{\rm m}$. To check the consistency of our
results we have performed a series of simulations for the \texttt{Soft}
EOS with low ($\Delta_\texttt{L} := 369\,{\rm m}$) and medium
($\Delta_\texttt{M} := 295\,{\rm m}$) resolution which is discussed in
Appendix \ref{sec:App1:conv}. For the spacetime evolution we use
\texttt{Antelope}~\citep{Most2019b}, which solves a constraint damping
formulation of the Z4 system~\citep{Bernuzzi:2009ex,Alic:2011a}.
Furthermore, to evolve the hydrodynamic part we use the
Frankfurt/Illinois (\texttt{FIL}) general-relativistic
magnetohydrodynamic code~\citep{Most2019b}, which is based on the
\texttt{IllinoisGRMHD} code \citep{Etienne2015}. \texttt{FIL} implements
fourth order conservative finite-differencing methods
\citep{DelZanna2007}, enabling a precise hydrodynamic evolution even at
low resolution. Furthermore, \texttt{FIL} is able to handle tabulated
EOSs that are dependent on temperature and electron-fraction. In
principle, \texttt{FIL} also includes a neutrino leakage scheme that can
handle neutrino cooling and weak interactions. However, in its current
form V-QCD does not include a description for neutrinos which is why we
do not make use of the leakage scheme in this work.

Finally, in Table~\ref{tab:NSprop} we summarise the properties of cold
non-spinning isolated neutron stars used in the construction of the
binary initial data as well as various characteristic frequencies
extracted from the post-merger waveforms presented in
Sec.~\ref{sec:gwspectra}.

\begin{figure*}[htb]
    \includegraphics[width=0.8\textwidth]{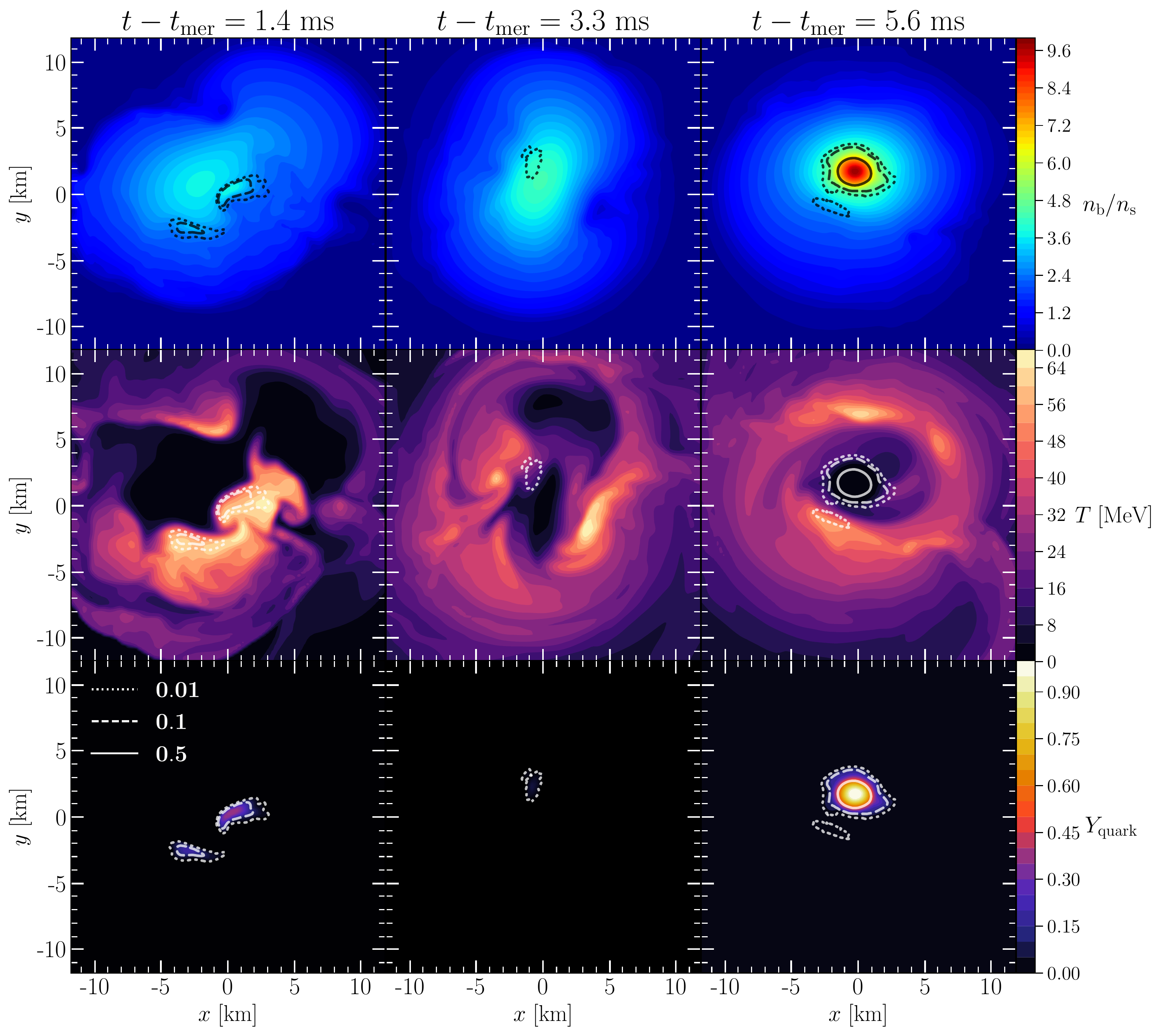}\quad
     \caption{ From top to bottom rows we show snapshots of the
       baryon-number density, temperature and quark fraction in the
       orbital plane at three different times $t-t_{\rm
         mer}=1.4,\,3.3,\,5.6\,\,{\rm ms}$ (from left to right) that are
       representative for HQ, WQ and CQ stages in the \texttt{Soft\_q07}
       simulation. In addition we mark contours for the quark fraction
       $Y_{\rm quark}=0.01,\,0.1,\,0.5$ by dotted, dashed and solid
       lines, respectively. }
    \label{fig:soft_q0.7_hr_3x3_plot2D}
\end{figure*}

\section{Merger Dynamics and Quark Formation}
\label{sec:dynamics}

\begin{figure*}[htb]
\includegraphics[width=0.99\textwidth, keepaspectratio]{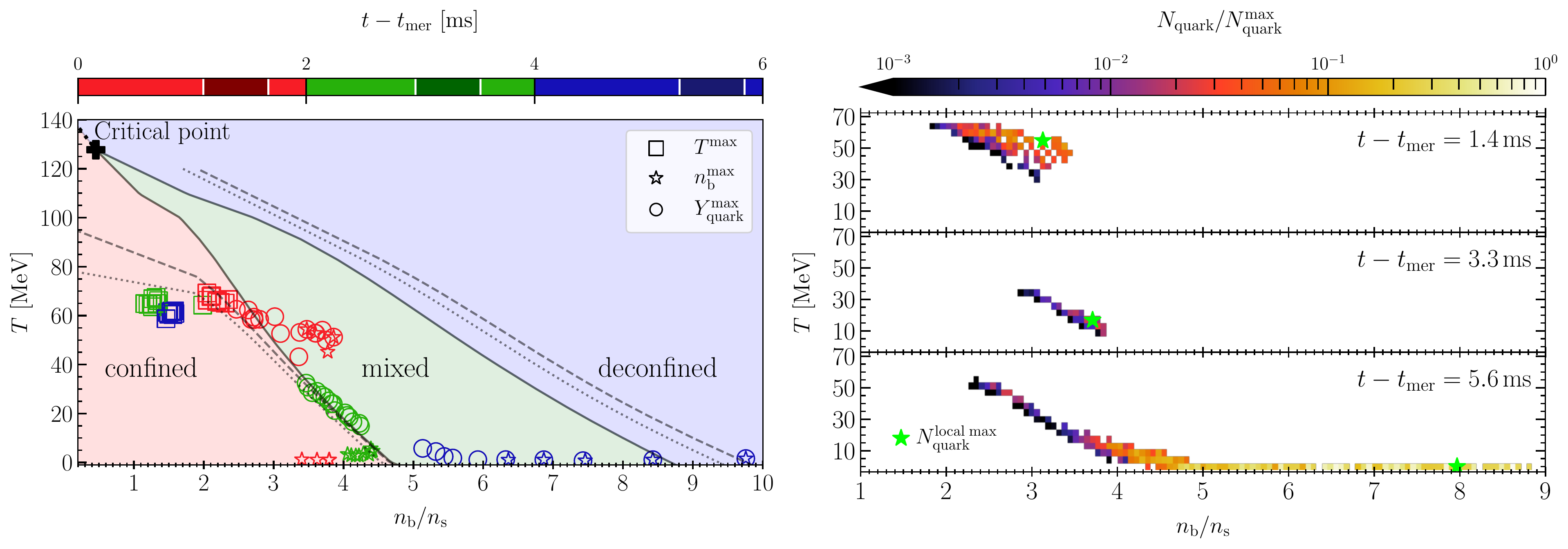}
     \caption{ Left: Phase diagram of the \texttt{Soft\_q07}
       post-merger evolution, with regions of confined, mixed and
       deconfined phases of matter shown. Points with a specific colour
       all occur within time intervals indicated by darker colour bars.
       Right: The amount of quarks in the orbital plane, measured using
       the same times as in Fig.~\ref{fig:soft_q0.7_hr_3x3_plot2D}. We
       normalize the measured amount of quarks by the maximal amount
       found in all three time slices presented here.
       }
    \label{fig:PhaseDiagram}
\end{figure*}

In this section we present the merger dynamics and discuss the three
mechanisms in which quarks are formed during the post-merger evolution.
We have performed numerical simulations of BNS systems using mass ratios
$q = 1 \,, 0.7$ where the neutron stars have an irrotational fluid
profile. For each configuration, the initial data and evolution have been
computed using the \soft, \inter, and \stiff~ EOSs in order to compare
the influence of stiffness on the quark production and lifetime of the
remnant HMNS whose lifetime is expected to be on the order of one
second~\cite{Gill2019,Murguia-Berthier2020}. A comparative overview of
the results from these numerical simulations can be seen in
Fig.~\ref{fig:yq_rho_temp_correl_annuli_combined} where we show the
evolution of the normalized maximum density, $(n_{\rm b}/n_{\rm s})_{\rm
  max}$, the maximum temperature, $T_{\rm max}$, and the quark fraction,
\Yq. To further characterise the distribution of quark matter within the
remnant HMNS, we define three regions that follow the centre-of-mass of
the HMNS in the orbital plane. We will refer to these regions from now on
as the inner core ($R<3\,{\rm km}$), outer core ($3\,{\rm
  km}$\,$<R<$\,$6\,{\rm km}$) and inner crust ($6\,{\rm
  km}$\,$<R<$\,$9\,{\rm km}$). An illustration along with further details
regarding the motivation behind these choices is discussed in Appendix
\ref{sec:App1:extraction}. The quark fraction $Y_{\rm quark}$ is
currently computed in the post-analysis of the evolutions. In order to
obtain the associated grid function values we use a multi-linear
interpolation scheme for the given tabulated EOS of a respective model,
in which discrete data points $Y_{\rm quark} = Y_{\rm quark}(n_{\rm b},
T, Y_{\rm e})$ are stored, as well as the 2D slices of the $n_{\rm b}$,
$T$ and $Y_{\rm e}$.

In all three models, comparing $\langle Y_{\rm quark}\rangle$ to the
maximal temperature shows a clear correlation between the appearance of
quark matter and the increase in temperature due to shock-heating during
the initial merger stage which is coloured red in
Fig.~\ref{fig:yq_rho_temp_correl_annuli_combined}. For the remainder of
this work, we will reference this as hot quark (HQ) production. The
amount of HQs produced is significantly larger for the \soft\ EOS than
for the stiffer models which is not unexpected as the densities inside a
neutron star with fixed mass is larger when matter is soft and is,
therefore, easier to produce strong shocks during the
collision. Furthermore, the amount of HQs produced is significantly
larger in the unequal mass case than in the equal mass case because the
initial density of the heavier companion is higher before merger in
unequal binaries which subsequently results in a more violent collision
and stronger shocks. Another significant difference between equal and
unequal mass binaries is that in the latter HQs are formed not only in
the inner core (and a tiny amount in the outer core), but also in
the outer core and inner crust region. This has to do with the fact that the
distribution of matter in the unequal mass case is highly
non-axisymmetric which leads to HQs in the hot off-central region of the
HMNS.

In the case of the \soft\ EOS, we see during the cooling of the remnant
and as the core becomes more dense an additional quark production channel
occurs. We will designate this production mechanism as warm quarks (WQ).
The formation of WQs appears to be a result of the complicated
post-merger dynamics that leads to a periodic expansion and contraction
of the HMNS. During the contraction the density increases close to the
rotation axis and leads to the formation of quarks. In the subsequent
expansion the density decreases and quarks transition back to baryonic
matter. In the \soft\ model we find that WQs are produced in the equal
mass case $\approx2\,{\rm ms}$ after the end of the HQ production, while
for unequal masses we find a continuous transition between the production
of HQs and WQs. For the stiffer models we do not find WQs, however, we
expect their formation to be possible with a heavier binary which will be
studied thoroughly in a future work. To capture accurately the delicate
combination of temperature and density required for WQs, the use of high
resolution numerical methods is essential. This is demonstrated in
Appendix \ref{sec:App1:conv} where we compare our results with two lower
resolutions for the \soft\ model and find that WQs can be absent in our
intermediate and low resolution simulations.

Finally, when the HMNS becomes approximately circular and its compactness
increases, a cold but dense quark matter core can be formed which we
denote as cold quark (CQ) production. We find CQ production to be
present only for the \soft\ model, which ultimately undergoes a
phase-transition-triggered collapse for the configurations
considered. Notice that due to the large latent heat in the V-QCD model,
no stable quark matter cores can be formed and therefore formation of a
quark core inevitably leads to the collapse of the remnant. The starting
time of CQ production depends on the mass ratio and the total mass of the
system. For the unequal-mass binary it starts already $\approx 4\,{\rm ms}$
after merger-time, while in the equal-mass case the starting time is
approximately two milliseconds later ($\approx 6\,{\rm ms}$). A close
inspection of the maximum number density in the core of the unequal-mass
binary reveals that CQs set in during the fifth bounce of the merger
remnant. In this stage the maximal temperature in the core is lower than
in the exterior, in particular for the equal-mass binary. We find this
stage to last for $\approx 4\,{\rm ms}$ ($\approx 2\,{\rm ms}$) for the
equal-(unequal-)mass binary, before the density and temperature in the
core eventually start to increase rapidly leading to the formation of a
black-hole horizon.

To further explore the appearance of HQ, WQ and CQ we show in
Fig.~\ref{fig:soft_q0.7_hr_3x3_plot2D} snapshots of the number density
(top), temperature (middle) and quark fraction (bottom) in the orbital
plane at three representative subsequent times (left to right) of the
post-merger evolution of the \texttt{Soft\_q07} configuration. The times
have been chosen based on the times of significant HQ, WQ, and CQ
production, respectively, as denoted by vertical dotted lines in
Fig.~\ref{fig:yq_rho_temp_correl_annuli_combined}. Additionally in
Fig.~\ref{fig:soft_q0.7_hr_3x3_plot2D} we indicate by dotted, dashed and
solid lines the outer contours of regions that contain quark fractions
larger than $0.01$, $0.1$ and $0.5$, respectively. The plots for
$t-t_{\rm mer}=1.4\,{\rm ms}$ clearly show the presence of HQs in the hottest
regions outside the dense core of the HMNS. Typical for this early post
merger state is the formation of multiple disconnected hot pockets with
temperatures well above $50$~MeV inside the HMNS which, in this case,
result in two disconnected regions where HQs are formed. The second
column shows snapshots at $t-t_{\rm mer}=3.3\,{\rm ms}$ corresponding to the
formation of a single patch of WQs. We note that a precise definition of
the WQ stage is difficult since it sensitively depends on the combination
of densities and temperatures outside the hottest and densest regions of
the star such that a transition to a mixed-phase is possible. These
conditions are typically realized in regions outside, but close to the
center of the HMNS where both temperature and density are significantly
below their maximal values. Finally, in the third column we show
snapshots at $t-t_{\rm mer}=5.6\,{\rm ms}$ where a pure quark matter core
($Y_{\rm quark}=1$) has already been formed and ultimately leads to a
phase-transition-triggered collapse of the HMNS. The maximal density of
the quark matter core is well above $9\,n_s$ (and rising), but the
temperature is very low further motivating our classification of quark
matter in this region as CQs. Also noteworthy is the appearance of a
small disconnected portion of WQs that forms in a small region outside
the dense and cold centre. It is interesting to compare our cold quark
core to~\cite{Weih:2019xvw}, who find the quark core to be hot. The
reason for this difference is that~\cite{Weih:2019xvw} models the
temperature dependence by adding a gamma-law to the cold EOS model where
the temperature simply scales with the density and does not take into
account the change in composition as in V-QCD.

To study the composition of matter in terms of the maxima of \Yq~
compared to $T_{\rm max}$ and $n_b^{\rm max}$, we show in
Fig.~\ref{fig:PhaseDiagram} (left) a phase diagram of the \soft\ V-QCD
model in the temperature and baryon-number density plane where red, green
and blue regions indicate confined, mixed and deconfined phases,
respectively; and solid lines mark the phase boundaries in beta
equilibrium which end at a critical point marked by a black cross.
Additionally, we show as a reference dashed and dotted lines indicating
the phase boundaries at approximately the minimal ($Y_e=0.05$) and
maximal ($Y_e=0.09$) values of the charge fraction inside the HMNS during
the evolution. These values differ from that of the beta equilibrium,
because we determine $Y_e$ by demanding beta equilibrium inside the stars
that are used to initialize the binary system, but advect $Y_e$ during
the subsequent evolution thereby driving the fluid out of beta
equilibrium. Finally, colored symbols mark representative results of the
$q=0.7$ simulation for the maximal temperature (squares), baryon-number
density (stars) and quark fraction (circles) collected during the early
HQ stage ($t-t_{\rm mer}=0-2\,{\rm ms}$, red), the WQ stage ($t-t_{\rm
  mer}=2-4\,{\rm ms}$, green) and the CQ stage ($t-t_{\rm mer}=4-6\,{\rm
  ms}$, blue). From the red symbols one clearly sees the maxima of the
quark fraction appearing close to the maxima in temperature at
approximately $60$\,MeV, while the maxima in the number density appear at
much lower temperature at the bottom of the phase diagram, illustrating
our motivation to classifying these quarks as HQs. The green symbols
represent examples for the WQ stage where $Y_{\rm quark}$ is maximal at
intermediate temperature ($\approx 20$\,MeV) and where the maximal
density ($\approx 4\,n_s$) occurs. The maximal temperature is
significantly higher ($\approx 60$\,MeV) in this case, but appears at
much lower densities ($\approx 1\,n_s$). Furthermore, the blue symbols
mark the CQ production where the maximal quark fraction coincides with
the maximum number density at very low temperature.

Finally, in Fig.~\ref{fig:PhaseDiagram} (on the right) we show our
measurements of the amount of quark matter during the three separate time-slices
used in Fig.~\ref{fig:soft_q0.7_hr_3x3_plot2D}.
For this we define $126\times 35$ bins with equal size in linear space, in the
$[n_s,\,10\, n_s]\times [0,140 \, \rm{MeV}]$ region of the number
density-temperature $(n_b, T)$ plane, which captures the whole domain of
quark production. To do the counting we evaluate the quantity $N_{\rm
  quark}=\sum_j V_j Y_{{\rm quark},j} n_{b,j}$ separately for each bin,
where the sum is performed over all the grid-cells $j$ of volume $V_j$ on
a fixed time-slice in the orbital plane, and then normalized by the
maximal count on the whole grid - where the maximal $N_{\rm
  quark}$ is marked with a star. The first row shows HQs at $t-t_{\rm
  mer}=1.4\,{\rm ms}$, which span temperatures $\approx40-60$\,MeV and
densities $\approx1-2\,n_s$. The second row shows WQs at $t-t_{\rm
  mer}=3.3\,{\rm ms}$ produced at temperatures $\approx10-30$\,MeV and
densities $\approx2-3\,n_s$. Finally, the third row for $t-t_{\rm
  mer}=5.6\,{\rm ms}$ clearly shows that significant amounts of WQs and
CQs can be present simultaneously. However, from the the colour code of
this plot one can also see that the majority of quarks (indicated in
yellow) are cold and dense.

\begin{figure*}[htb]
    \includegraphics[width=0.88\textwidth, keepaspectratio]{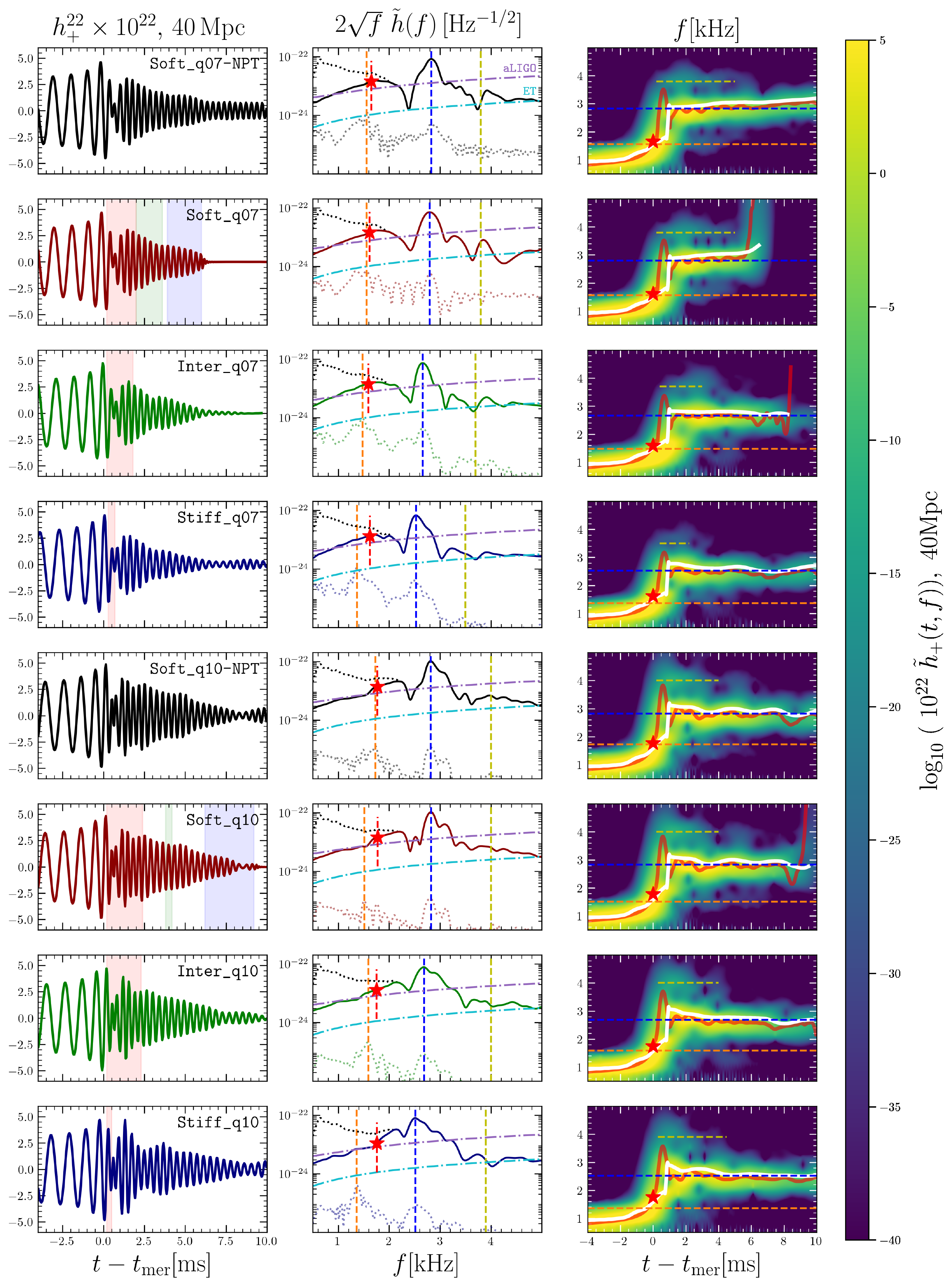}\quad
     \caption{ Shown above is the $h^{22}_+$ of the strain, the 
       power-spectral density of the effective strain, and the spectrogram of
       the $h^{22}_+$ of the \soft, \inter, and the \stiff~ EOSs
       considered in this work for $q = 1 \,, 0.7$ and $\mathcal{M}_{\rm
         chirp} = 1.188$. Also included are the related spectra for
       \softnpt~ which does not include a phase transition to quark
       matter. In the right two panels the dashed blue and orange lines
       correspond to the $f^{2,2}$ and $f^{2,1}$ peaks respectively, the
       dashed yellow lines corresponds to the $f_3$ peak as measured in
       the spectrogram, and the star denotes the peak merger frequency, $f_{\rm
         mer}$. In the right figure, the white line traces
       the maximum in the spectrogram. Finally, the sensitivity curves in
       the middle plot are related the current sensitivity of advanced
       LIGO and the Einstein Telescope respectively
       \cite{adLIGO2018,Punturo:2010zza}.
    }
    \label{fig:gwtable}
\end{figure*}

\section{Gravitational Wave Analysis}
\label{sec:gwspectra}

In this section we analyse the gravitational waveforms and their spectral
properties. We use the Newman-Penrose formalism
\cite{Alcubierre:2006,Bishop2016} to relate the Weyl curvature scalar
$\psi_4$ to the second time derivative of the polarization amplitudes of
the gravitational wave strain $h_{+,\times}$ via
\begin{equation}\label{eq:psi4}
  \ddot{h}_++i\ddot{h}_\times=\psi_4
  :=\sum_{\ell=2}^{\infty}\sum_{m=-\ell}^{m=\ell}\psi_4^{\ell,m}
     {_{-2}Y}_{\ell,m}\,,
\end{equation}
where $_sY_{\ell,m}(\theta,\phi)$ are spin-weighted spherical harmonics
of weight $s=-2$. From our simulations we extract the modes
$\psi_4^{\ell,m}$ with a sampling rate of $\approx 26\,{\rm kHz}$ from a
spherical surface with $\approx 440\,{\rm km}$ radius centred at the
origin of our computational domain and extrapolate the result to the
estimated luminosity distance of $40\,{\rm Mpc}$ of the GW170817 event
\cite{LIGOScientific:2017vwq}. In addition, we fix the angular dependence
of the spherical harmonics part by the viewing angle $\theta=15^{\circ}$
determined from the jet of GW170817~\cite{Ghirlanda:2018uyx} and set
$\phi=0^{\circ}$ without loss of generality. We restrict our analysis to
the dominant $\ell=m=2$ and sub-dominant $\ell=2,\, m=1$ modes of the
expansion \eqref{eq:psi4}. To analyze the spectral features of the
waveforms we follow~\cite{Takami:2014zpa} and compute the power spectral
density (PSD) defined as
\begin{align}
  \label{eq:PSD}
  \tilde{h}^{\ell,m}(f):=\frac{1}{\sqrt{2}}
  \biggl(
  &\left|\!\int\! \dd t\,{\rm e}^{-2\pi ift}
     h^{\ell,m}_+(t)\right|^2 + \nonumber\\
  &   \left|\!\int\! \dd t\,{\rm e}^{-2\pi ift}
    h^{\ell,m}_\times(t)\right|^2
  \biggr)^{1/2}\,,
\end{align}
where the time integration is performed over the interval $t-t_{\rm
  mer}=0-10\,{\rm ms}$ or up to the collapse to a black hole as is the
case for the \soft\ binaries. Furthermore, we use spectrograms to study
the time dependence of the gravitational wave frequency distribution. To
perform the Fourier transform for the spectrograms we use time-windows of
$3\,{\rm ms}$ centred at every $\approx0.04\,{\rm ms}$ of our waveform
data. We also study the phase difference between the $h_+^{2,2}$ and
$h_\times^{2,2}$ modes by computing the instantaneous gravitational wave
frequency $f_{\rm GW}$ defined as
\begin{equation}\label{eq:fGW}
    f_{\rm GW}:=\frac{1}{2 \pi} \frac{d \phi}{dt}\,,\qquad \phi:={\rm
      arctan} \left( \frac{h_\times^{2,2}}{h_+^{2,2}} \right)\,.
\end{equation}
Finally, we can measure from $f_{\rm GW}(t)$ the instantaneous frequency
at merger time defined as $f_{\rm mer}=f_{\rm GW}(t=0)$.

In Fig.~\ref{fig:gwtable} we include a summary of our gravitational wave
analysis of the configurations described in Tab.~\ref{tab:NSprop}. From
top to bottom we show results for the different EOS models and the two
mass ratios we have simulated. In the first column we show the dominant
$h_{+}^{2,2}$ gravitational wave strain component with the time periods
where the dominant quark production mechanism, HQ, WQ and CQ, are
highlighted in red, green and blue, respectively.

The most prominent signature of quark matter in the waveform is clearly
the earlier termination of the signal due to a phase-transition-triggered
collapse to a black hole. This happens in our simulations only for the
\soft\ model, which for both mass ratios forms a CQ stage that leads to a
collapse of the HMNS. We note that models that include
hyperons~\cite{Sekiguchi:2011mc,Radice:2016rys} can lead to signatures
similar as those from a first order deconfinement phase transition
\cite{Bauswein:2018bma}. In both cases a softening of the EOS leads to
more compact merger remnants that is closer to collapse. We estimate the
lifetime $t_{\rm BH}$ of our merger remnants by the time at which the
strain amplitude starts to decay exponentially as characteristic for the
black-hole ring-down. In cases where no horizon is formed during
simulation time, we provide lower limits on $t_{\rm BH}$ by the final
time of the simulation. The values for $t_{\rm BH}$ we obtain in this
way are listed in Table~\ref{tab:NSprop}. Comparing the waveforms of the
\soft\ model for $q=0.7$ (second row) and $q=1$ (sixth row) shows that
the unequal-mass case collapses $\approx 4\,{\rm ms}$ earlier than the equal
mass case. This behaviour is consistent with the threshold mass
analysis~\cite{Tootle:2021umi,Kashyap:2021wzs,Koelsch2021}, which finds
that highly asymmetric mergers tend to collapse at lower total mass than
symmetric ones.

In the second column of Fig.~\ref{fig:gwtable} we show the PSDs computed
via Eq.~\eqref{eq:PSD}. Solid lines represent the dominant $\ell=m=2$
mode, while light dotted lines are the sub-dominant $\ell=2,\, m=1$
mode. The dotted black line is the contribution to the PSD that comes
from the inspiral part ($t-t_{\rm mer}<0$), which is not included in the
other curves. In addition we show estimated sensitivities of the advanced
LIGO~\cite{adLIGO2018} (purple dot-dashed) and the Einstein
Telescope~\cite{Punturo:2010zza} (cyan dot-dashed line). The PSD shows
the typical three-peak structure from which a number of characteristic
post merger frequencies can be extracted. The most prominent peak
$f^{2,2}$ (dashed blue) corresponds to the dominant post-merger frequency
of the $\ell=m=2$ mode of the signal. Comparing the three EOS models for
$q=1,0.7$ we find good agreement amongst their values of $f^{2,2}$ and
only a small shift to lower frequencies with increasing stiffness of the
EOS. Most notably, the $f^{2,2}$ peaks for the \texttt{Soft\_q10} and
\texttt{Soft\_q10-NPT} are equivalent though the peak-width for
\texttt{Soft\_q10-NPT} is slightly broader (where the \texttt{NPT} 
simulations do not include a transition to quark matter 
as we will discuss below). In addition, we mark the
maximal frequency $f^{2,1}$ of the sub-dominant $\ell=2, m=1$ mode by a
dashed orange line and the merger frequency $f_{\rm mer}$ by a red star.
Similar to the dominant mode, the $f^{2,1}$ frequencies (dashed orange)
decrease only slightly with the increase in stiffness, but there is an
appreciable difference when measuring the $f^{2,1}$ for the \texttt{Soft}
EOS with and without a phase transition to quark matter in the equal-mass
binaries. Overall we find that the sub-dominant mode in the equal mass
case is by at least one order of magnitude suppressed with respect to the
unequal-mass case. This is expected, since unequal-mass binaries result
in less axially symmetric remnants than equal-mass binaries and,
therefore, produce a richer post-merger frequency spectrum. In this sense
gravitational waves from asymmetric binaries potentially provide more
information about the EOS than symmetric binaries. However, in both cases
the amplitude of the sub-dominant mode is orders of magnitude too low to
be seen by current detectors, even when assuming the advantageous
luminosity distance and sky localization of GW170817.

In the third column of Fig.~\ref{fig:gwtable} we show spectrograms where
the colours from dark blue to light yellow indicate the distribution of
low and high spectral weight during the evolution. In addition, we show
lines for various characteristic frequencies: dashed blue and orange
lines mark the$f^{2,2}$ and $f^{2,1}$ respectively - as extracted from the
PSD. The dashed yellow line marks $f_3$ which we extract directly from
the spectrogram because it is difficult to obtain unambiguously from the
PSD. Finally, the solid red line is $f_{\rm GW}$ as computed from
\eqref{eq:fGW} and the solid white line follows the maximum values of the
spectrogram. Note that the values of $f^{2,2}$ and $f^{2,1}$ extracted
from the PSD align exceptionally well with the $f_1$ and $f_2$ peaks seen
in the spectrogram; however, the $\ell,m = (2,2),(2,1)$ PSDs do not
provide identifying information related to the $f_3$ peak which is
measured directly from the spectrogram.

\subsection{Collapse Behaviour and Lifetime}

To isolate the impact of quark matter on the lifetime we also perform
simulations where we exclude the thermodynamically preferred quark matter
phase at high temperature and densities in the \soft\ V-QCD model 
(see first and fifth row in Fig.~\ref{fig:gwtable}) 
which we designate as the \softnpt~model. Unlike the case
with quark matter, the equal-mass case without quark matter does not
collapse during our simulation time, however, we find that in the
unequal-mass case, a binary using \softnpt~model collapses after $\approx
11\,{\rm ms}$, i.e.\,, roughly $5\,{\rm ms}$ later than with the \soft\ model. The
early collapse of our simulations with the \soft\ V-QCD model point to a
strong tension with the expected second-long lifetime of GW170817
\cite{Gill2019,Murguia-Berthier2020}. This means our results disfavour
the \soft\ EOS and, therefore, allow us to further constrain the V-QCD
model with information gained from the post-merger phase. This is an
important result since, as discussed in Sec.~\ref{sec:Model}, the cold
part of the \soft\ V-QCD model is fully consistent with currently known
observations and the constraints from the post-merger lifetime can only
be inferred from merger simulations. 

Furthermore, obtaining faithful estimates for the lifetime of merger
remnants that survive significantly longer than $\approx10\,{\rm ms}$ is
difficult in our setup. There are mainly two reasons for this: First of
all, resolving the formation of quark matter requires relatively fine
resolution, which makes long simulation times numerically extremely
expensive. We therefore restrict our high-resolution simulations to
$\approx 10\,{\rm ms}$ after merger. However, we present in
Appendix~\ref{sec:App1:medruns} simulation results with medium resolution
up to $> 35\,{\rm ms}$ for the \inter~ and \stiff~ model. This allows us to
approximate the lower bound of their lifetime to be $>35\,{\rm ms}$. Second of
all, the competition between turbulent heating in the presence of
magnetic fields and cooling effects by neutrino emission can become
important on time-scales on the order of several $100\,{\rm ms}$
\cite{Fujibayashi:2020dvr}. Since we do not include either of these
effects in our simulations we expect the thermodynamic properties, in
particular the temperature distribution inside the star, not to be
reliable on such long time-scales. However, we expect these thermal
effects to be negligible on the comparably short time-scales the
life-time estimates listed in Table~\ref{tab:NSprop} are based on.

\section{Summary and Conclusion\label{sec:Conclusion}}

We have studied fully 3+1 dimensional GRHD simulations of GW170817-like
events using a unified model for QCD that includes the deconfinement
phase transition from baryonic to quark matter. The focus of our work
was the formation of quark matter during and after the merger of the two
neutron stars. For this we studied three variants of the unified model,
namely a soft, an intermediate and a stiff version. We identified three
distinct mechanisms in the post-merger evolution during which mixed
baryonic and quark matter and pure quark matter is formed and which we
denote as: i) hot quark (HQ), ii) warm quark (WQ) and iii) cold quark
(CQ) production. We find that some or all of these stages may be present
in the evolution depending on the masses of the stars and the EOS model
employed.

Specifically, the HQ production appears at merger time and we find it to
last for approximately two milliseconds, except for the stiff model where
only a small amount of quarks are produced at merger time. During this
time, strong shocks that are formed during the initial contact of the two
stars lead to a steep rise in temperature. This causes a local formation
of quark matter in the hottest, but not in the dense regions of the
merger remnant. As the remnant begins to cool and lose its differential
rotation causing an increase in density, the production of WQs may be
possible depending on the model and binary configuration. The WQ
production is the result of complicated merger dynamics and is observed
in regions that are typically neither the hottest nor the densest regions
of the HMNS. In this work WQs are only observed in the soft model. It is
important to note that the existence and starting time of the WQ stage
depends strongly on the details of the complicated post-merger evolution
during which the HMNS pulsates in a highly non-axisymmetric way and, as such,
turns out to be particularly sensitive to the mass ratio of the
binary system. Additionally, we find that WQs are produced earlier and
also in a larger quantities in non-equal mass binaries than in equal-mass
binaries. Finally, the CQ production appears after the most violent
post-merger period has settled and the density in the centre increases
and cools down. The production of CQ is then formed in the dense and
cold centre of the merger remnant. In this work, the CQ stage is
followed by a phase-transition-triggered collapse to a black hole, using
the terminology of~\cite{Weih2020}. The reason for this is the large
latent heat of the deconfinement phase transition in the V-QCD model at
low temperatures which does not allow stable cold quark matter cores
inside isolated stars or HMNSs.

Finally, we studied the waveforms and their spectral properties and find
the most significant imprint is the early termination of the
gravitational wave signal due to the phase-transition-triggered collapse
due to CQ production. By this we find the short post-merger lifetime of
$\lesssim 10\,{\rm ms}$ obtained from the soft version of V-QCD to be in
tension with the expected second long lifetime of GW170817. It is
remarkable that, even though the majority of the post-merger period
($\sim 6ms$) of the unequal-mass case for the soft model includes a phase
transition to a mixed or pure quark phase, there is essentially no
measurable imprint on the amplitude or frequency of the gravitational
wave strain.
Moreover, this tension also means that there are new constraints for the
predictions for the QCD equation of state arising from the merger
simulations. As the soft version of the unified model is basically
excluded by the lifetime estimate of the remnant, the soft versions
should be also excluded from the analysis of the
cold~\cite{Jokela:2020piw,Jokela:2021vwy} and hot~\cite{Demircik:2021zll}
equation of state. We do not attempt to make such constraints precise in
this article, but as an example we discuss the location of the critical
point of the nuclear to quark matter transition. It was found
in~\cite{Demircik:2021zll} that the temperature $T_c$ of the critical
point varies roughly between 110~MeV and 130~MeV, with the soft version
predicting the highest temperature. Exclusion of the soft model therefore
means that $T_c \lesssim 120$~MeV within the framework.

There are multiple directions we plan to explore in future work. In the
present work we only considered non-spinning binaries with individual
masses consistent with GW170817, which prefers the low-spin prior. It
will be important to investigate the dependence of various quark matter
production stages also on different individual masses and spins of the binary
components. Another direction worth studying is the impact of the
deconfinement phase transition on the threshold mass, which has been
analysed extensively in~\cite{Tootle:2021umi,Kashyap:2021wzs,Koelsch2021,Bauswein2020c}
for purely hadronic EOSs as well as some hybrid models that include first
order phase transitions\cite{Kashyap:2021wzs,Koelsch2021,Bauswein2020c}.
Notice also that our simulations did not take into account the surface
tension at the interface between the nuclear and quark matter. The value
of the surface tension has not yet been computed in the V-QCD framework,
and even if the value was known, it is not straightforward to take it
into account: strong first order phase transitions in the presence of a
sizeable surface tension proceed through bubble nucleation, which is a
dynamical process that cannot be easily taken into account at the level
of the equation of state. It is however clear that the impact of the
surface tension would be to somewhat reduce the amount of mixed phase and
suppress the quark production. We plan to study in a future work how large
this effect is.
Furthermore, it would be interesting to include cooling effects from
neutrino emission which could have non-trivial impact on the quark
formation, especially during the HQ and WQ stages. Finally, because of
the high computational costs we restricted our high-resolution
(medium-resolution) simulations to $\approx 10\,{\rm ms}$ ($\approx
40\,{\rm ms}$) after the merger. This allowed us to obtain reliable
predictions for the quark matter production and the lifetime from the
simulations that collapse during this short post-merger period. However,
we can only speculate about the ultimate fate of the cases that do not
collapse during our simulation time. We plan to extend these simulation
to the order of $1$~second by imposing axial symmetry in future long-term
evolutions to obtain improved estimates for the lifetime of our
long-lived HMNSs.

\begin{acknowledgments}

CE acknowledges support by the Deutsche Forschungsgemeinschaft (DFG,
German Research Foundation) through the CRC-TR 211 'Strong-interaction
matter under extreme conditions'-- project number 315477589 -- TRR 211.
ST, KT, and LR acknowledge the support by the State of Hesse within the
Research Cluster ELEMENTS (Project ID 500/10.006). TD and MJ have been
supported by an appointment to the JRG Program at the APCTP through the
Science and Technology Promotion Fund and Lottery Fund of the Korean
Government. TD and MJ have also been supported by the Korean Local
Governments -- Gyeong\-sang\-buk-do Province and Pohang City -- and by
the National Research Foundation of Korea (NRF) funded by the Korean
government (MSIT) (grant number 2021R1A2C1010834). LR acknowledges
funding by the ERC Advanced Grant ``JETSET: Launching, propagation and
emission of relativistic jets from binary mergers and across mass
scales'' (Grant No. 884631). The simulations were performed on HPE Apollo
HAWK at the High Performance Computing Center Stuttgart (HLRS) under the
grant BNSMIC. For our visualisations we made an extensive use of the
Kuibit library \cite{kuibit21}.
\end{acknowledgments}

\appendix
\begin{figure*}[htb]
  \includegraphics[width=0.35\textwidth,keepaspectratio]{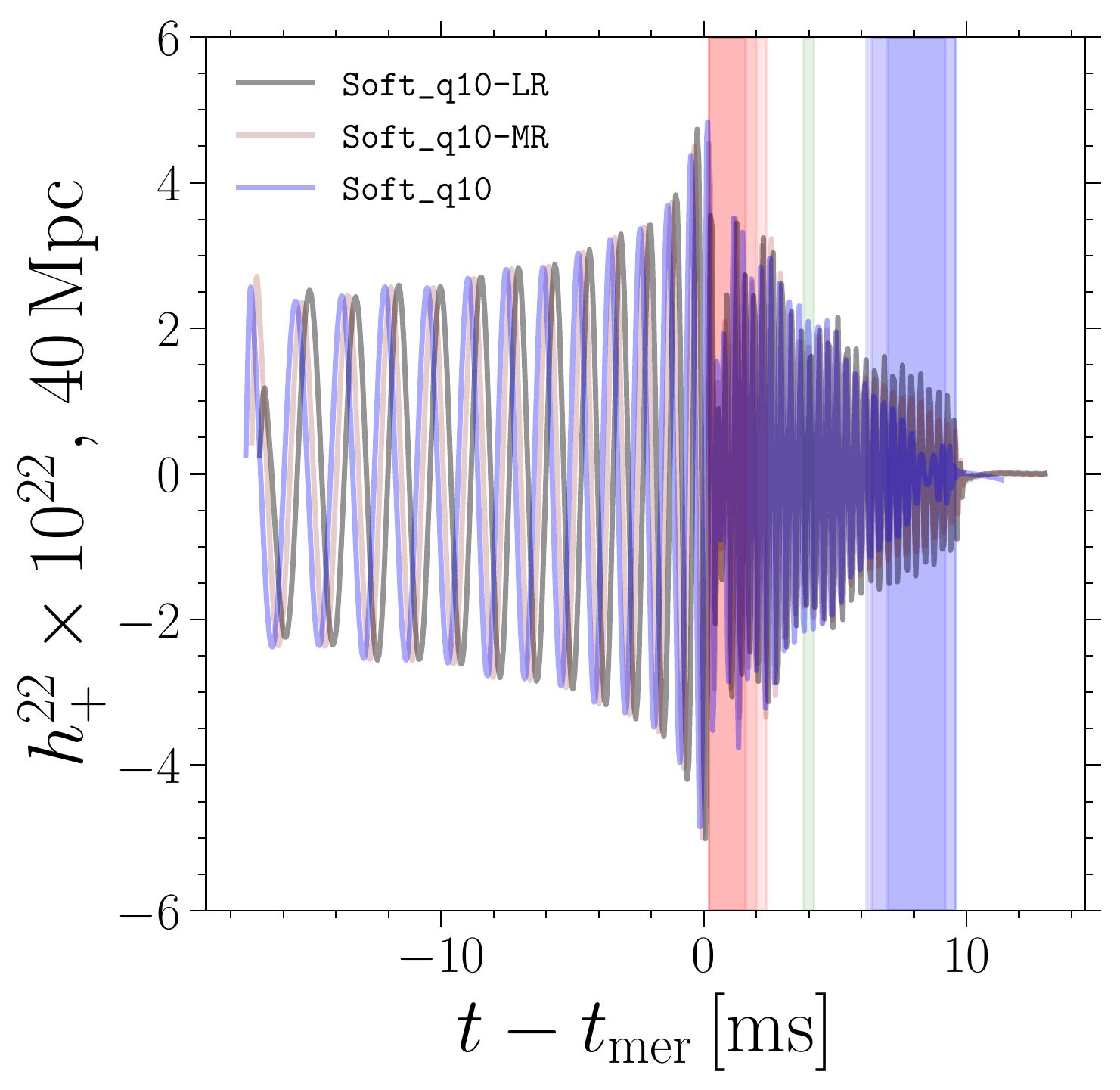}
  \hskip 0.5cm
  \includegraphics[width=0.53\textwidth,keepaspectratio]{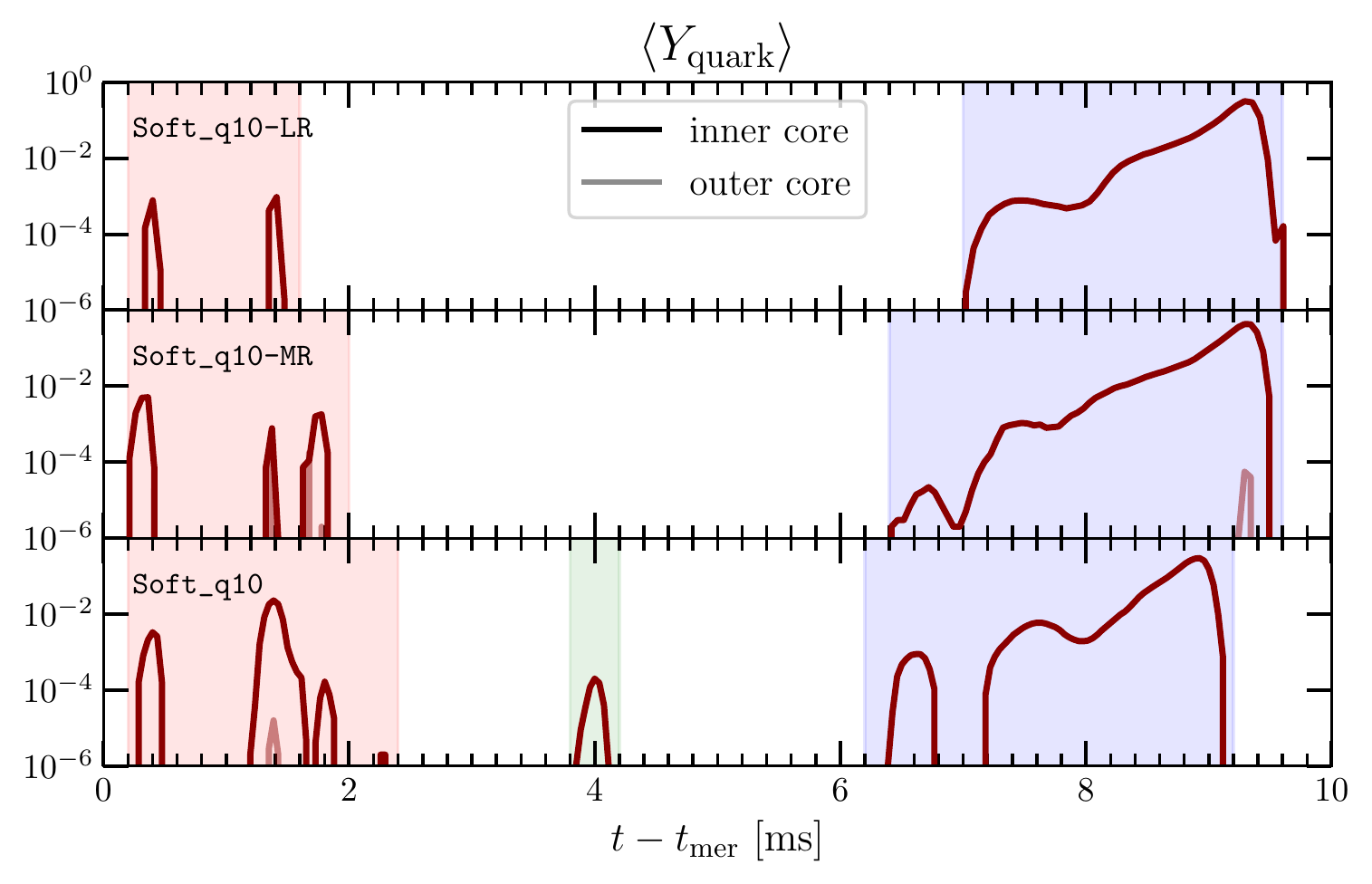}
     \caption{Shown are the consistency results of the \texttt{Soft\_q10}
       dataset. Left: consistency as measured in the gravitational wave strain $h^{22}_+$
       polarisation amplitude. Although slight changes in the amplitude
       and phase are observed, the collapse time is the same in all
       resolutions. Right: the corresponding quark fraction for the three
       resolutions considered. Most notable is the appearance of
       WQ in the high-resolution run further emphasising the
       need for high resolutions to capture fine-structure details. }
    \label{fig:convergence}
\end{figure*}

\section{Convergence Analysis}
\label{sec:App1:conv}

To ensure a high level of confidence in the discussed results, we have
performed a consistency study with the \texttt{Soft\_q10} initial data to
ensure that the overall dynamics of the merger and the phase
transition-induced collapse are not simply artefacts of the numerical
accuracy. To this end, we have employed three resolutions which are
characterised by the finest grid spacing of $\Delta_\texttt{L}
:=369\,{\rm m}$, $\Delta_\texttt{M} :=295\,{\rm m}$ and
$\Delta_\texttt{H} :=221\,{\rm m}$ where $\Delta_\texttt{H}$ is the
resolution used for the results presented in the main body of the
text. Additionally, we employ a Courant static time step such that $dt =
C_{dt} {\rm max}(dx)$ where $C_{dt} = 0.2$ for our
simulations. Therefore, an increase in spatial resolution results in an
increase in temporal resolution.

As shown in Fig. \ref{fig:convergence} (left) the gravitational waveforms
show only a slight variation in the relative phase during the inspiral
and a systematic decrease in the post-merger amplitude, however, all
resolutions considered result in the same collapse time. Furthermore, we
see in Fig. \ref{fig:convergence} (right) that the existence and
identification of the different stages of quark production is robust
under the change of resolution. Notably, the intermediate production of
WQ appears for $\Delta_\texttt{H}$ which is not surprisingly
absent in the $\Delta_\texttt{M}$ resolution given the sensitivity
required in temperature and density for WQ to appear especially
for such a short duration ($\sim 0.3\,{\rm ms}$). Remarkably, the peaks in
$Y_{\rm quark}$ follow a similar spacing in time, where individual peaks
tend to form more often with an increase in resolution.

\section{Medium Resolution Runs}
\label{sec:App1:medruns}

\begin{figure*}[htb]
    \includegraphics[width=1\textwidth,keepaspectratio]{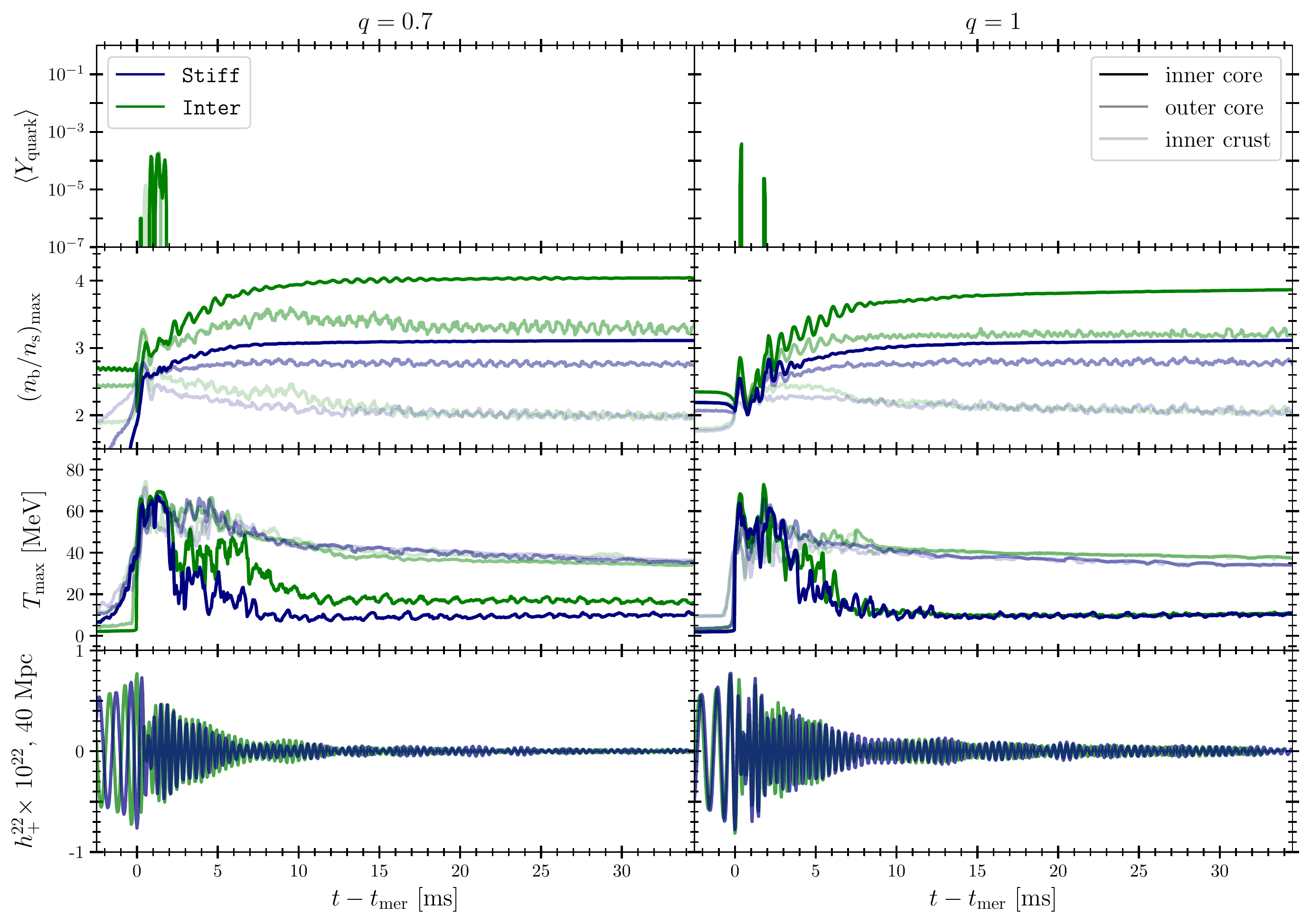}
    \caption{The results of the \texttt{Stiff} and \texttt{Inter} medium
      resolution runs for both investigated mass ratios. In neither do we
      observe a collapse throughout the whole $\sim 35$ ms of the
      post-merger evolution. }
    \label{fig:mediumRes}
\end{figure*}

Given the consistent nature of our evolutions as discussed in
\ref{sec:App1:conv}, we conducted additional evolutions at
$\Delta_\texttt{M}$ resolution up to $\approx 35\,{\rm ms}$ for the
\inter~ and \stiff~ models. The results of these runs are summarised in
Fig.~\ref{fig:mediumRes} where the behaviour of the thermodynamic
quantities is shown to be consistent with the results in
Fig. \ref{fig:yq_rho_temp_correl_annuli_combined} as well as the
waveforms shown in Fig. \ref{fig:gwtable}. This allows us to place lower
bounds on the collapse time, namely $\approx 35\,{\rm ms}$. Although
these results provide evidence that the \inter~ and \stiff~ models are
consistent with the expected second long lifetime of GW170817, we want to
emphasize that our simulations do not include contributions of magnetic
fields and neutrino cooling which could be relevant for the long-term
stability of a HMNS.

\begin{figure}[htb]
  \includegraphics[width=0.45\textwidth,keepaspectratio]{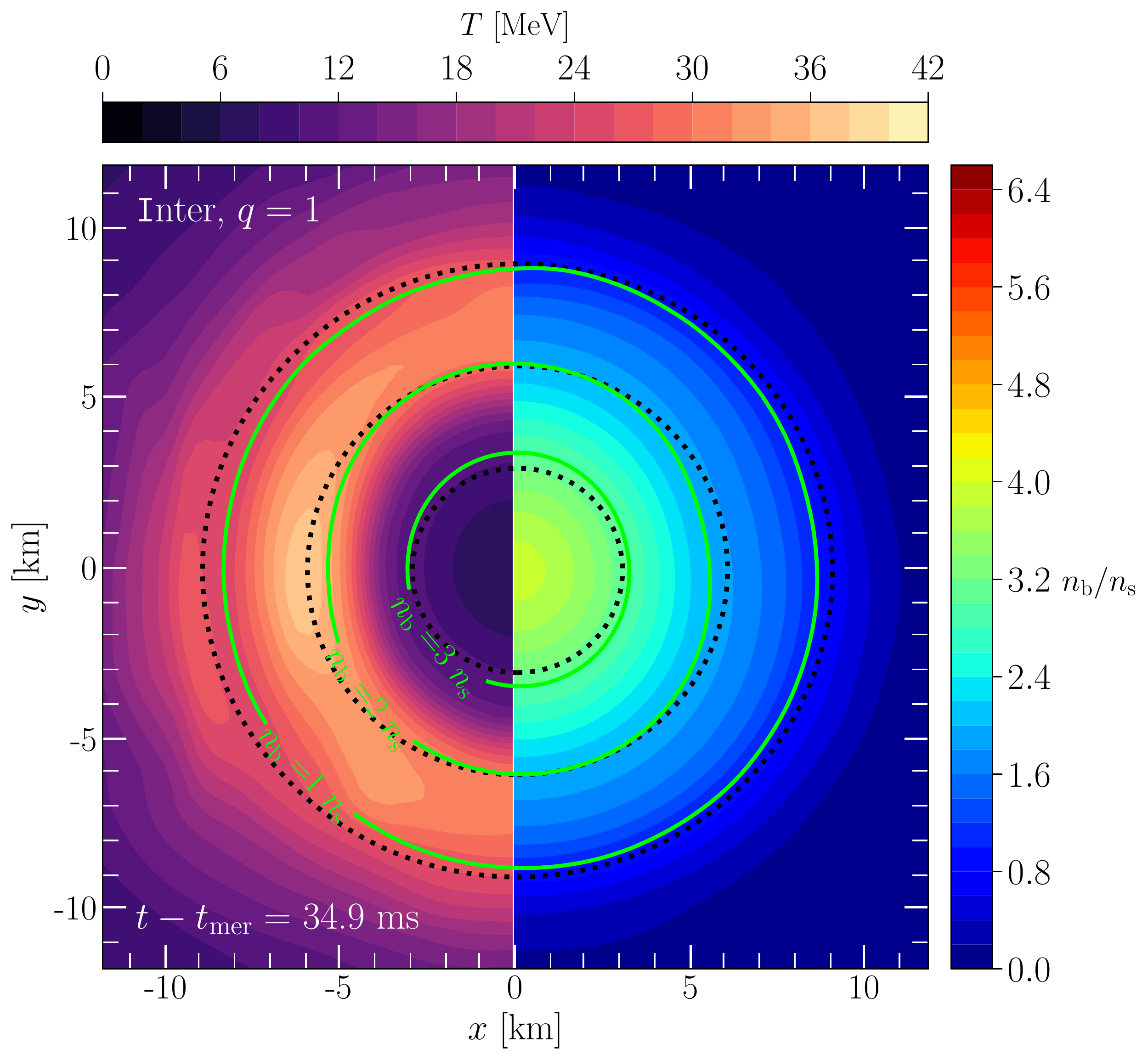}
  \caption{A slice on $(x,y)$ plane from the \texttt{Inter\_q10-MR} run,
    plotted at the latest available time. Shown are the temperature (left
    part) and the baryon-number density (right part), while the black
    dashed lines represent the $3, 6$, and $9\,{\rm km}$ ranges over
    which the integrals are computed. Note the good match with the
    normalised baryon-number $n_{\rm b}/n_s$ at values of $3, 2$, and
    $1$, respectively (green solid isocontours).}
    \label{fig:radii}
\end{figure}

\section{Extraction Radii of Thermodynamic Quantities}
\label{sec:App1:extraction}

In our analysis of the quark production within the remnant HMNS it seemed
pertinent to not only discuss the amount of quarks produced, but also the
relative location of quark production. To illustrate the extraction radii
we present in Fig.~\ref{fig:radii} these regions for the
\texttt{Inter\_q10} run, for which there is no collapse to a black hole
on the timescale that we cover. Lime contours within the figure
correspond to baryon-number density $n_{\rm b}$ equal to $1n_{\rm s}$,
$2n_{\rm s}$ and $3n_{\rm s}$. The regions enclosed by the black dotted
circles correspond to radii of $3$ km, $6$ km and $9$ km which we
designate as the inner core, outer core and inner crust. The annuli are
centred on the maximal density point for $t<t_{\rm mer}$ and on the
centre of mass of the system for $t \geq t_{\rm mer}$. These extraction
domains are then used to extract the local average and maximal values of
thermodynamic quantities as shown in
Fig.~\ref{fig:yq_rho_temp_correl_annuli_combined} and
Fig.~\ref{fig:mediumRes}. The averages over any annuli $\Omega$ are
obtained using the following formula:
\begin{equation}
\label{eq:integralsformula}
\langle f \rangle := \frac{1}{A} \int_{\Omega} f d^{2}x \,, \qquad A :=
\int_{\Omega} d^{2}x\,,
\end{equation}
where, for simplicity, we consider the metric determinant of flat
spacetime in the integrand.

\bibliography{aeireferences}
\end{document}